\documentclass[10pt]{article}

\usepackage{amsmath,amssymb,amsthm}
\usepackage{amsxtra}
\usepackage{amsfonts}
\usepackage{amssymb}
\usepackage{url}
\usepackage{hyperref}
\usepackage{esint}
\usepackage{eucal}
\usepackage{mathrsfs}
\usepackage{cite}
\usepackage{graphicx,graphics}
\newcommand\fig[1] {{\rm Figure}~\ref{fig:#1}}
\newcommand\labfig[1] {\label{fig:#1}}
\newcommand\sect[1] {\ref{sect:#1}}
\newcommand\labsect[1] {\label{sect:#1}}

\newcommand\eq[1] {(\ref{#1})}
\newcommand{\bfm}[1]{\mbox{\boldmath ${#1}$}}

\newcommand{\nonum}{\nonumber \\}
\newcommand{\beqa}{\begin{eqnarray}}
\newcommand{\eeqa}[1]{\label{#1}\end{eqnarray}}
\newcommand{\beq}{\begin{equation}}
\newcommand{\eeq}[1]{\label{#1}\end{equation}}

\newcommand{\Real}{\mathop{\rm Re}\nolimits}

\newcommand{\Ga}{\alpha}
\newcommand{\Gb}{\beta}

\newcommand{\Ge}{\epsilon}

\newcommand{\Gg}{\gamma}
\newcommand{\Gc}{\chi}

\newcommand{\Gs}{\sigma}

\newcommand{\Go}{\omega}

\newcommand{\GO}{\Omega}

\newcommand{\BGG}{\bfm\Gamma}
\newcommand{\BGL}{\bfm\Lambda}

\newcommand{\BGY}{\bfm\Psi}

\newcommand{\CE}{{\cal E}}

\newcommand{\CH}{{\cal H}}

\newcommand{\CJ}{{\cal J}}

\newcommand{\CT}{{\cal T}}
\newcommand{\CU}{{\cal U}}

\newcommand{\bpm}{\begin{pmatrix}}
\newcommand{\epm}{\end{pmatrix}}

\def\Ba{{\bf a}}

\def\Bh{{\bf h}}

\def\Bk{{\bf k}}

\def\Bp{{\bf p}}
\def\Bq{{\bf q}}

\def\Bs{{\bf s}}

\def\Bw{{\bf w}}
\def\Bx{{\bf x}}

\def\BA{{\bf A}}
\def\BB{{\bf B}}
\def\BC{{\bf C}}

\def\BE{{\bf E}}
\def\BF{{\bf F}}
\def\BG{{\bf G}}

\def\BI{{\bf I}}
\def\BJ{{\bf J}}
\def\BK{{\bf K}}
\def\BL{{\bf L}}
\def\BM{{\bf M}}
 
\def\BO{{\bf O}}
\def\BP{{\bf P}}

\def\BR{{\bf R}}
\def\BS{{\bf S}}
\def\BT{{\bf T}}

\def\BW{{\bf W}}

\usepackage{graphics}
\usepackage{amssymb}

\title{
    A unifying perspective on linear continuum equations prevalent in science. Part VI: rapidly converging series expansions for their solution}
\author{}
\date{}
\begin{document}
\maketitle
\vskip -.5cm
\centerline{\large Graeme W. Milton}
\centerline{Department of Mathe A unifying perspective on linear continuum equations prevalent in science. Part VI: rapidly converging series expansions for their solutionmatics, University of Utah, USA -- milton@math.utah.edu.}
\vskip 1.cm
\begin{abstract}
  We obtain rapidly convergent series expansions of resolvents of operators taking the form $\BA=\BGG_1\BB\BGG_1$
  where $\BGG_1(\Bk)$ is a projection that acts locally in Fourier space and $\BB(\Bx)$ is an
  operator that acts locally in real space. Such resolvents arise naturally when one wants
  to solve any of the large class of linear physical equations surveyed in Parts I, II, III, and IV
  that can be reformulated as problems in the extended abstract theory of composites.
  We show how the information about the spectrum of $\BA$ can be used to
  greatly improve the convergence rate.
\end{abstract}
\section{Introduction}
\setcounter{equation}{0}
\labsect{30}
In Parts I, II, III, and IV
\cite{Milton:2020:UPLI, Milton:2020:UPLII, Milton:2020:UPLIII, Milton:2020:UPLIV}
we established that an avalanche of equations in science can be rewritten in the form
\beq \BJ(\Bx,t)=\BL(\Bx,t)\BE(\Bx,t)-\Bs(\Bx,t),\quad \BGG_1\BE=\BE,\quad\BGG_1\BJ=0,
\eeq{ad1}
as encountered in the extended abstract theory of composites, $\BGG_1=\BGG(\Bk)$ is a projection
operator that acts locally in Fourier space, and $\Bs(\Bx)$ is the source term. In
Part V \cite{Milton:2020:UPLV} we established the connection between solving these
equations and computing resolvents of operators of the form $\BA=\BGG_1\BB\BGG_1$
where $\BB=\BB(\Bx)$ acts locally in real space.

Here in Part VI we are concerned with using rapidly converging series expansion
for the solution of \eq{ad1} to obtain rapidly converging series expansions for
resolvents of the form
\beq \BR_0=(z_0\BI-\BA)^{-1}=z_0(\BI-\BA/z_0)^{-1}, \eeq{0.1}
where the operator $\BA$ takes the form $\BA=\BGG_1\BB\BGG_1$, in which
$\BB=\BB(\Bx)$ acts locally in real space and typically has an inverse,
and one that is easily computed. 
Thus if $\BGG_1$ or $\BB$ act on a field $\BF$ to produce a field $\BG$ then we have, respectively,
that $\BG(\Bx)=\BB(\Bx)\BF(\Bx)$ or $\widehat{\BG}(\Bk)=\BGG_1(\Bk)\widehat{\BF}(\Bk)$, in
which $\widehat{\BG}(\Bk)$ and $\widehat{\BF}(\Bk)$ are the Fourier components of $\BG$ and $\BF$.

As in the previous parts
we define the inner product of two fields $\BP_1(\Bx)$ and $\BP_2(\Bx)$ to be
\beq (\BP_1,\BP_2)=\int_{\mathbb{R}^3}(\BP_1(\Bx),\BP_2(\Bx))_{\CT}\,d\Bx,
\eeq{innp}
where $(\cdot,\cdot)_{\CT}$ is a suitable inner product on the space $\CT$
such that the projection $\BGG_1$ is selfadjoint with
respect to this inner product, and thus the space $\CE$ onto which
$\BGG_1$ projects is orthogonal to the space $\CJ$ onto which
$\BGG_2=\BI-\BGG_1$ projects. We define the norm of a field $\BP$ to
be $|\BP|=(\BP,\BP)^{1/2}$, and given any operator $\BO$ we define its norm
to be
\beq \|\BO\|=\sup_{\BP,\,|\BP|=1}|\BO\BP|. \eeq{on}
When we have periodic fields in periodic media the integral in \eq{innp} should be taken over the unit cell $\GO$ of periodicity.
If the fields depend on time $t$ then we should set $x_4=t$ take the integral over $\mathbb{R}^4$ with the integral over the spatial variables
restricted to $\GO$ if the material and fields are spatially periodic. 

The goal of this paper is to review iterative methods that have been developed to accelerate the solution of problems in the extended theory
of composites, and to transfer this knowledge to develop rapidly convergent
iterative schemes for the calculation of resolvents, where $\BA=\BGG_1\BB\BGG_1$. These iterative methods automatically apply to
calculating the action of the inverse of a matrix $\BB$ on a vector subspace, when the inverse of $\BB$
on the whole vector space is easily computed. They were first introduced by Moulinec and Suquet \cite{Moulinec:1994:FNM} in the context of calculating the fields and
effective moduli in the theory of composites, and subsequently accelerated algorithms were discovered: see \cite{Eyre:1999:FNS, Michel:2000:CMB, Monchiet:2011:PBF} and Chapter 8 of \cite{Milton:2016:ETC}.
They have been the subject of increasing attention: see \cite{Zhou:2020:OSA} and references therein.

The work presented is largely based on the articles
\cite{Moulinec:1994:FNM, Eyre:1999:FNS, Moulinec:2018:CIM, Milton:2019:NRF}
  and Chapter 8 of \cite{Milton:2016:ETC}, but develops some of the ideas further.

  Solutions to the equations \eq{ad1} in the extended abstract theory of composites
  are easily expressed in terms of the related resolvent
  \beqa \BR & = &  \BR_0+(\BGG_1-\BI)/z_0
  =(z_0\BI-\BGG_1\BB)^{-1}\BGG_1=(z_0\BGG_1-\BA)^{-1} \nonum
& = & (\BGG_1\BL\BGG_1)^{-1}\quad \text{where}\quad \BL=z_0\BI-\BB,
\eeqa{0.1a}
where in the last two expressions for $\BR$,
the inverse is to be taken on the subspace onto which  $\BGG_1$ projects: $\BR$ is the
resolvent of $\BA$ within this subspace.

We seek expansions such that the
action of the resolvent on a given field can be calculated by a simple iterative process, that for a given $z_0$ just requires the
application of a given operator to the previous iterate. This avoids having to store multiple fields, such as the $m$ different fields
that result from the actions of the $m$ operators $\BA, \BA^2, \ldots, \BA^m$ on the given field. Even if we are interested in
$\BR$ as a function of $z_0$ the rapid convergence implies that, to achieve a desired accuracy, we can keep fewer fields than if the convergence were slower.
One reason for the importance of knowing the resolvent as a function of $z$ is that it allows computation of any operator
valued analytic function $f(\BA)$ of the matrix $\BA$ according to
the formula
\beq f(\BA)=\frac{1}{2\pi i}\int_\Gg f(z_0)(z_0\BI-\BA)^{-1}\,dz_0,
\eeq{-0.1d}
where $\Gg$ is a closed contour in the complex plane that encloses the spectrum of $\BA$.

The first equation in \eq{ad1} is called the constitutive law with $\Bs(\Bx)$ being the source term. As remarked previously, if the null space of $\BL$ is nonzero
then one may one can often shift $\BL(\Bx)$ by a
multiple $c$ of a ``null-$\BT$ operator''
$\BT_{nl}(\Bx)$ (acting locally in real space or spacetime,
and discussed further in Section 3 of \cite{Milton:2020:UPLV}), defined to have the
property that
\beq \BGG_1\BT_{nl}\BGG_1=0, \eeq{nl-1}
that then has an associated quadratic form (possibly zero) that is
a ``null-Lagrangian''.
Clearly the equations \eq{ad1} still hold, with $\BE(\Bx)$ unchanged
and $\BJ(\Bx)$ replaced by $\BJ(\Bx)+c\BT_{nl}\BE(\Bx)$
if we replace $\BL(\Bx)$ with  $\BL(\Bx)+c\BT_{nl}(\Bx)$. In other
cases $\BL$ may contain $\infty$ (or $\infty$'s) on its diagonal.
If one can remove any degeneracy of $\BL(\Bx)$, we can consider the
dual problem
\beq \BE=\BL^{-1}\BJ(\Bx)+\BL^{-1}\Bs(\Bx),\quad \BGG_2\BJ=\BJ,
\quad\BGG_2\BE=0,
\eeq{dual}
with $\BGG_2=\BI-\BGG_1$, and then, if desired, try to shift  $\BL^{-1}(\Bx)$
by a multiple
of a ``null-$\BT$ operator'' $\widetilde{\BT}_{nl}(\Bx)$ satisfying
$\BGG_2\widetilde{\BT}_{nl}\BGG_2=0$ to remove its degeneracy. 

Our results, in particular, apply to the family of problems associated with analyzing the response of two phase composite materials, 
where $\BB(\Bx)$ itself depends on $z_0$ and takes the form
\beq \BB(\Bx)=z_0\BI-\BL_1\Gc_1(\Bx)-\BL_2\Gc_2(\Bx), \eeq{0.2}
where the $\Gc_i(\Bx)$ are the characteristic functions
\beqa\Gc_i(\Bx) &= & 1\quad \text{in phase}\,\,i \nonum
&= & 0\quad \text{elsewhere},
\eeqa{0.3}
satisfying $\Gc_1(\Bx)+\Gc_2(\Bx)=1$, while $\BL_1$ and $\BL_2$ are the tensors of the two phases, representing their material properties, and
the ``reference parameter'' $z_0$ can be freely chosen. This family will serve as model problems for our analysis. Specifically,
the convergence of the expansions that we develop is best illustrated if we further assume that
\beq \BB(\Bx)=z_0\BI-z_1\BI\Gc_1(\Bx)-z_2\BI\Gc_2(\Bx)=(z_0-z_2)\BI-(z_1-z_2)\BI\Gc_1(\Bx), \eeq{0.4}
where now, for example, $z_1$ and $z_2$ may represent the conductivities of the two phases and $z_0$ a reference conductivity. With the particular choice
$z_0=z_2$ the expression \eq{0.1} reduces to
\beq \BR=z_2^{-1}\{\BI-(1-z_1/z_2)\BGG_1\Gc_1\BGG_1\}^{-1},
\eeq{0.5}
which is now again a problem directly of the form \eq{0.1} with $\BB$ and $z_0$ now being identified as
\beq \BB=\Gc_1\BI,\quad z_0=z_2/(z_2-z_1). \eeq{0.5a}

We will assume that $z_0$, $\BGG_1$ and $\BB$ are fixed and known. So the analysis in this paper is really just about computing
the inverse of operators of the form $\BI-\BGG_1\BB\BGG_1/z_0$. The parameter $z_0$, even if fixed, is helpful as the rates of convergence
of the series we investigate are conveniently expressed in terms of  $z_0$.


\section{Some elementary series expansions}
\setcounter{equation}{0}
\labsect{31}
We start by assuming that $\BB$ is real and that we know some bounds on it:
\beq b^-\BI\leq \BB \leq b^+\BI,\quad \text{implying}\quad b^-\BGG_1\leq \BA \leq b^+\BI, \eeq{0.7}
where the last identity follows by projecting the first inequality on the subspace $\CE$. 
We may sometimes know tighter bounds on $\BA$:
\beq a^-\BI\leq \BA \leq a^+\BGG_1,\quad \text{where}\quad a^-\geq b^-,\quad a^+\leq b^+. \eeq{0.7a}
Some approaches to deriving such bounds have been given in Section 3 of \cite{Milton:2020:UPLV}.

One well known expansion of the resolvent is the Laurent series:
\beq \BR(z_0)/z_0=(\BI-\BA/z_0)^{-1}=\sum_{n=0}^\infty (\BA/z_0)^{n},
\eeq{0.6}
better known as the Neumann expansion or Born expansion in the context of operators $\BA$,
which holds provided the series converges and this is the case if the matrix or operator $\BA/z_0$ has norm less than $1$. From
the bounds \eq{0.7a} it follows that
\beq |\BA/z_0|\leq r_0,\quad\text{where}\quad r_0=\frac{\max\{a^+,a^-\}}{|z_0|}\leq \frac{\max\{b^+,b^-\}}{|z_0|},
\eeq{0.8}
and convergence of the expansion is assured if $r_0<1$, i.e. for $|z_0|>\max\{a^+,a^-\}$. With $\BB$ and $z_0$ being given by \eq{0.5a}
we can take $b^-=0$ and $b^+=1$ and \eq{0.6} naturally reduces to
\beq  \BR=z_2\sum_{n=0}^\infty (1-z_1/z_2)^n(\BGG_1\Gc_1)^{n}.
\eeq{0.9}
 
As shown for example in Section 2 of \cite{Milton:2020:UPLV} of Part V, the solution of \eq{ad1} is $\BE=\BR\Bs$ where
$\BR$ can be expressed in various equivalent forms including
\beq \BR=[\BGG_1\BL\BGG_1]^{-1}=[\BI-\BGG\BB]^{-1}\BGG=
\BL^{-1}-\BL^{-1}[\BI-\widetilde{\BGG}\widetilde{\BB}]^{-1}\widetilde{\BGG}\BL^{-1}
\eeq{RR1}
where
\beq \BB(\Bx)=\BL_0-\BL(\Bx),\quad  \widetilde{\BB}(\Bx)=\BM_0-\BL^{-1}(\Bx)
\eeq{RR2}
are operators that are local in real space, in which $\BL_0$ and $\BM_0$ are constant
reference tensors, and where
\beq \BGG=\BGG_1(\BGG_1\BL_0\BGG_1)^{-1}\BGG_1,\quad
\widetilde{\BGG}=\BGG_2(\BGG_2\BM_0\BGG_2)^{-1}\BGG_2
\eeq{RR3}
act locally in Fourier space, the inverses being respectively on the spaces $\CE$ and $\CJ$
onto which $\BGG_1$ and $\BGG_2=\BI-\BGG_1$ project.
With $\BL_0=z_0\BI$, we have that $\BGG=\BGG_1/z_0$ and then it is apparent that with $\BA=\BGG_1\BB\BGG_1$ where $\BB=z_0\BI-\BL$,
$\BR$ given by \eq{RR1} is in fact the resolvent \eq{0.1a} when we consider $\BB$ to
be fixed and $\BL$ to be a function of $z_0$. Conversely, if we are
interested in computing the resolvent in \eq{0.1}, or equivalently \eq{0.1a},
then we can recast it as a problem in the theory of composites with $\BL=z_0\BI-\BB$.
Having established this connection with the resolvent we can now apply all the theory developed in extended abstract theory
of composites to resolvents of the required form, and conversely.

For sufficiently small $\BB$ we get the series expansion
\beq \BR=[\BGG_1\BL\BGG_1]^{-1}=[\BI-\BGG\BB]^{-1}\BGG=\sum_{n=0}^\infty [\BGG\BB]^{n}\BGG=\sum_{n=0}^\infty [\BGG(\BL_0-\BL)]^{n}\BGG.
\eeq{1.12}
Although for fixed $\BL$ each term in this series depends on $\BL_0$ the sum is independent of $\BL_0$ when the series converges. The choice of $\BL_0$ influences the rate
of convergence, and indeed whether the series converges or not.
The series expansion \eq{1.12} is well known in the theory of composites:
see, for example Chapter 14 of \cite{Milton:2002:TOC},
\cite{Willis:1981:VRM}, and references therein.

Alternatively, if
$\widetilde{\BB}$ is sufficiently small we have the expansion 
\beq [\BI-\widetilde{\BGG}\widetilde{\BB}]^{-1}\widetilde{\BGG}=
\sum_{n=0}^\infty [\widetilde{\BGG}\widetilde{\BB}]^n\widetilde{\BGG},
\eeq{1.12alt}
which may be inserted in \eq{RR1} to get a different series expansion for $\BR$.

For the special case of a two phase medium where $\BB(\Bx)$ takes the
form \eq{0.2} we may take $\BL_0=\BL_2$ giving
$\BB(\Bx)=\Gc(\Bx)(\BL_1-\BL_1)$ and obtain the expansion
\beq \BR=[\BI-\BGG\Gc(\BL_2-\BL_1)]^{-1}\BGG
=\sum_{n=0}^\infty [\BGG\Gc(\BL_2-\BL_1)]^{n}\BGG,
\eeq{2ph}
that is convergent for $\BL_1$ that is sufficiently close to $\BL_2$. More precisely, if $\BL_2$
is real and positive definite, then using that $\Gc$ and $(\BL_2)^{1/2}\BGG(\BL_2)^{1/2}$
are selfadjoint projections, we have
\beq \|\BGG\Gc(\Bx)(\BL_2-\BL_1)\|
=\|(\BL_2)^{1/2}\BGG(\BL_2)^{1/2}\Gc[\BI-(\BL_2)^{-1/2}\BL_1(\BL_2)^{-1/2}]\|
\leq \|\BI-(\BL_2)^{-1/2}\BL_1(\BL_2)^{-1/2}\|,
\eeq{2pha}
so the series converges if $\|\BI-(\BL_2)^{-1/2}\BL_1(\BL_2)^{-1/2}\|<1$.

 We can now take rapidly converging iterative methods for the solution of
 \eq{ad1} and apply them to obtain rapidly convergent series expansions for the resolvent. These expansions, that will be a major focus of the paper,
 give the action of the resolvent on a source field $\Bs$ in the form
  \beq \BR\Bs=\BC_0\sum_{j=0}^\infty \BW^j\Bs,
 \eeq{it1}
 for suitable operators $\BC_0$ and $\BW$, whose action is relatively easy to compute (typically just requiring two fast Fourier transforms: to Fourier space and back).
 The iterative procedure of obtaining the fields
 \beq \Bq_{i+1}=\BW\Bq_{i}+\Bs,\quad \Bq_0=\Bs, \eeq{it2}
 gives $\BC_0\Bq_{n}$ as a good approximation to $\BR\Bs$ for large enough $n$. There is no need to keep the $\Bq_j$, for $j\leq i$ once one has computed $\Bq_{i+1}$.
 If one has a series expansion of the form
 \beq \BR\Bs=\BC_0\sum_{j=0}^\infty c^j \BW^j\Bs,
 \eeq{1t3}
 and one is interested in $\BR\Bs$ as a function of $c$ (which may in turn
 be a function of another variable of interest, such as $z_0$), then one can replace the iterative procedure in \eq{it2} with
 \beq \Bq_{i+1}=\BW\Bq_{i},\quad \Bq_0=\Bs, \eeq{1t4}
 storing the $\Bq_{i}$ as one goes along. Then if the series converges rapidly, the approximation
 \beq   \BR\Bs \approx\BC_0\sum_{j=0}^n c^j\Bq_{j} \eeq{it5}
 holds for relatively small values of $n$. Of course as $c$ is increased the series converges more slowly, or perhaps not at all,
 and then the approximation becomes poor for small values of $n$
 
\section{Improvements to the Neumann or Born Series}
\setcounter{equation}{0}
\labsect{32}
We start by reviewing a well known route for improving the convergence rate of the Neumann or Born Series, that does
not rely on the fact that we can express $\BA$ in the form $\BA=\BGG_1\BB\BGG_1$. Thus, we note that $\BA$ can be
split as $\BA=(\BA-c\BI)+c\BI$ and the resolvent can be re-expressed as
\beq \BR_0=[(z_0-c)\BI-(\BA-c\BI)]^{-1}=(z_0-c)^{-1}\{\BI-[(\BA-c\BI)/(z_0-c)]\}^{-1},
\eeq{0.9a}
where $c$ can be chosen to make the associated expansion
\beq \BR=(z_0-c)^{-1}\sum_{n=0}^\infty(\BA-c\BI)^n/(z_0-c)^n \eeq{0.12}
converge more rapidly than the expansion with $c=0$: the basic idea here is to choose $c$ to shift $\BA$ to decrease the spectral radius.
Such splittings are well known for accelerating convergence, the best known being the Jacobi and Gauss-Seidel splittings \cite{Golub:1996:MC}.
The expansion can clearly be calculated iteratively. If $\BB$ satisfies the bound \eq{0.7} then a natural choice of $c$ is
\beq c=\tfrac{1}{2}(b^++b^-), \eeq{0.12a}
giving
\beq |\BA-c\BGG_1|\leq \Ga,\quad\text{where}\quad \Ga=\tfrac{1}{2}(b^--b^-),
\eeq{0.12b}
and the series \eq{0.12} is guaranteed to converge if
\beq |(\BA-c\BI)|/|(z_0-c)|\leq|(\BB-c\BGG_1)/(z_0-c)|\leq r_1,
\eeq{0.12c}
where
\beq r_1=\left|\frac{b^+-b^-}{b^+ +b^- -2z_0}\right|
=\left|\frac{(b^+-z_0)-(b^--z_0)}{(b^+-z_0)+(b^--z_0)}\right|=\left|\frac{q-1}{q+1}\right|\quad\text{in which}\quad q= (z_0-b^+)/(z_0-b^-).
\eeq{0.12d}
Improved convergence can be obtained if we have bounds on $\BA$ itself that are tighter than the bounds \eq{0.7}.

We now draw upon rapidly converging iterative methods for the solution of
\eq{ad1} in the extended theory of composites and apply them to obtain rapidly convergent series expansions for the resolvent.
This will be the focus of the rest of the paper.

In particular, as \eq{RR1} holds for any choice of $\BL_0$  we can transform to an equivalent problem where $\BB$ is replaced with
\beq \BB'=\BL_0'-\BL=\BB-\BL_0+\BL_0',
\eeq{2.1}
and we have the identity
\beq [\BI-\BGG\BB]^{-1}\BGG=[\BI-\BGG'\BB']^{-1}\BGG' \quad \text{with} \quad
\BGG'=\BGG_1(\BGG_1\BL_0'\BGG_1)^{-1}\BGG_1,
\eeq{2.2}
and $\BGG$ being given by \eq{RR3}. The associated series expansion when
$\BL_0=z_0\BI$ is
\beq [\BI-\BGG_1\BB/z_0]^{-1}\BGG
=\sum_{n=0}^\infty [\BGG'(\BB-z_0\BI+\BL_0')]^{n}\BGG'.
\eeq{2.2a}

Let us suppose that $\BB$ is Hermitian and satisfies the bound \eq{0.7}. We take
$\BL_0'=z_0'\BI$ with
\beq z_0'=z_0-c,\quad \text{where}\quad c=\tfrac{1}{2}(b^+ +b^-),
\eeq{2.3}
so that $\BB'=\BB-c\BI$ satisfies
\beq     -\tfrac{1}{2}(b^+-b^-)\leq \BB'\leq \tfrac{1}{2}(b^+-b^-). \eeq{2.4}
This implies $|\BB'/z_0'|\leq r_1$ where
\beq r_1=\left|\frac{b^+-b^-}{b^+ +b^- -2z_0}\right|
=\left|\frac{(b^+-z_0)-(b^--z_0)}{(b^+-z_0)+(b^--z_0)}\right|=\left|\frac{q-1}{q+1}\right|,\quad\text{in which}\quad q= (z_0-b^+)/(z_0-b^-),
\eeq{2.5}
and the series expansion
\beq [\BI-\BGG\BB]^{-1}\BGG
=  [\BI-\BGG'\BB']^{-1}\BGG'
=\sum_{j=0}^\infty [\BGG'\BB']^{j}\BGG'
=\sum_{n=0}^\infty [\BGG_1\BB'/z_0']^{n}\BGG_1/z_0'
\eeq{2.6}
will converge provided $r_1<1$, i.e. provided $z_0<b^-$ or $z_0>b^+$. Moreover $r_1$ determines the minimum rate of convergence.
In the field of composites the series expansion \eq{1.12} and the independence of the resulting sum on $\BL_0$ (assuming the sum of the series converges)
is well known. Moulinec and Suquet \cite{Moulinec:1994:FNM} realized that the series could be easily computed by an iterative process as in \eq{it2}.
The action of $\BB$ (or $\BB'$) can be
computed in real space while the action of $\BGG$ (or $\BGG'$) can be computed in Fourier space and Fast Fourier transforms can be used to transform
between real space and Fourier space. The choice \eq{2.3} is motivated by their choice of a ``reference medium''. Moreover, and importantly, their approach is easily
extended to nonlinear media \cite{Moulinec:1994:FNM}, and has successfully been used for studying elastoplasticity, elastoviscoplasticity, dislocations,
shape memory polycrystals, and crack prediction in brittle materials: See \cite{Zhou:2020:OSA} and references therein, where
Zhou and Bhattacharya use a related  augmented Lagrangian method, also introduced in the
accelerated scheme of Michel, Moulinec, and Suquet \cite{Michel:2000:CMB}, to study bifurcations and liquid crystal elastomers.
 
By substituting the formula $\BB'=\BB-c\BI$ in \eq{2.6} with $c=\tfrac{1}{2}(b^++b^-)$
we obtain
\beq  [\BI-\BGG\BB]^{-1}\BGG=\sum_{n=0}^\infty [\BGG_1\BB'/z_0']^{n}\BGG_1 =\sum_{i=0}^\infty (\BA-c\BI)^n/(z_0-c)^{n}\BGG_1,
\eeq{2.7}
which is exactly the same expansion as in \eq{0.12} in view of the identity \eq{0.1a}.
The advantage of the expansion \eq{2.2a} is that it allows more general
choices of $\BL_0'$ not necessarily proportional to $\BI$.

In particular, for the resolvent \eq{0.5} with $\BB$ and $z_0$ being given by \eq{0.5a} so that $b^+=1$, $b^-=0$ and $c=1/2$ we obtain
$z_0'=\tfrac{1}{2}(z_2+z_1)/(z_2-z_1)$ and the expansion \eq{2.6} becomes
\beq
\BR=z_2\sum_{n=0}^\infty [\BGG_1(\Gc_1-\tfrac{1}{2}\BI)]^{n}/(z_0')^n=z_2\sum_{n=0}^\infty r^n[\BGG_1(2\Gc_1-\BI)]^{n},\quad
\text{where}\quad r=\frac{z_2-z_1}{z_2+z_1}
\eeq{2.8}
Comparing this with \eq{0.9} we now have an expansion where $\Gc_1$ is replaced by $\Gc_1-\tfrac{1}{2}\BI$ which has half the spectral radius.

The expansion
still converges for an appropriate value of $z_0'$ when $\BB$ is not
Hermitian, but, for some $z_0$, $\BL=\BL_0-\BB=z_0\BI-\BB$ is 
bounded and coercive in the sense that there is some $\Ga>0$ and $\Gb>0$ such that
\beq \Gb>\|\BL\|,\quad\Real(\BL\BP,\BP)>\Ga|\BP|^2,\quad\text{for all}\quad \BP.
\eeq{coerc}
Then, as proved in Section 2.4 of \cite{Milton:2016:ETC},
with $z_0'=\Gb^2/\Ga$ one gets the bound
\beq \|\BB'/z_0'\|=\|\BI-\BL/z_0'\|\leq\sqrt{1-(\Ga/\Gb)^2}<1,
\eeq{convest}
which ensures convergence of the series \eq{2.6}.

\section{An accelerated convergence method}
\setcounter{equation}{0}
\labsect{33}
To obtain, in most cases,  accelerated convergence we use the identity
\beqa [\BI-\BGG_1\BB/z_0]^{-1}=[\BI-\BGG'(\BL_0'-\BL)]^{-1}
&=&[\BI+\BM(\BL-\BL_0')-(\BM-\BGG')(\BL-\BL_0')]^{-1}\nonumber\\
&=&[\BI+\BM(\BL-\BL_0')]^{-1}[\BI-(\BM-\BGG')(\BL-\BL_0')[\BI+\BM(\BL-\BL_0')]^{-1}]^{-1} \nonumber\\
&=&(\BL-\BL_0')^{-1}\BK(\BI-\BGY\BK)^{-1},
\eeqa{3.1}
where
\beq \BK=(\BL-\BL_0')[\BI+\BM(\BL-\BL_0')]^{-1},\quad \BGY=\BM-\BGG'.
\eeq{3.2}
This identity had its genesis in formulae for the fields and effective tensors in laminated materials \cite{Milton:1990:CSP},
later independently arrived at in \cite{Zhikov:1991:EHM}. Then it was further employed in representations for the effective conductivity
of a composite as a function of the component conductivities: see equation (5.20) in \cite{Milton:1990:RCF}. It was used in \cite{Eyre:1999:FNS} to develop the fast
numerical schemes that we generalize here (see also sections 14.9, 14.10, and 14.11 in \cite{Milton:2002:TOC}). It also
has proved invaluable for the development of the theory of exact relations in composites \cite{Grabovsky:1998:EREa, Grabovsky:2000:ERE}
(see also Chapter 17 in \cite{Milton:2002:TOC} and the book \cite{Grabovsky:2016:CMM}), and in the affiliated development of
exact identities satisfied by the Green's function (fundamental solution) in certain classes of inhomogeneous media (not
necessarily with microstructure) and the
associated discovery of a wealth of new conservation laws, called boundary field equalities \cite{Milton:2019:NRF}.

The rate of convergence of the series is enhanced when $\BM$ is chosen to make the norm of $\BGY$ small. When $\BL_0$ is positive
definite we have that $\BGG\leq\BL_0^{-1}$ and so a natural choice is $\BM=\tfrac{1}{2}\BL_0^{-1}$. In this case
\beq \BK=2(\BL-\BL_0')(\BL+\BL_0')^{-1}\BL_0',\quad \BGY=(\BL_0')^{-1}(\BI-2\BGG'\BL_0')/2.
\eeq{3.3}
Further let us suppose that $\BL_0'=z_0'\BI$. Then we obtain
\beq |(\BL-\BL_0')(\BL+\BL_0')^{-1}|= |(\BL-z_0'\BI)(\BL+z_0'\BI)^{-1}|\leq r_2,
\eeq{3.4}
where
\beq r_2= \max\left\{\frac{|z_0-b^+-z_0'|}{|z_0-b^++z_0'|}, \frac{|z_0-b^--z_0'|}{|z_0-b^-+z_0'|}  \right\}.
\eeq{3.5}
We choose
\beq z_0'=\sqrt{(z_0-b^+)(z_0-b^-)}
\eeq{3.6}
to minimize $r_2$, giving
\beq r_2=\frac{\sqrt{q} -1}{\sqrt{q} +1},
\eeq{3.7}
where $q= (z_0-b^+)/(z_0-b^-)$ is the same as that given in \eq{2.5}.
The value of $q$ is always greater than $1$ when the series converges, i.e. provided $z_0<b^-$ or $z_0>b^+$.
Comparing this with the expression for $r_1$, we see that we get faster convergence since $\sqrt{q}$ is
smaller than $q$ and significantly smaller when $q$ is large. 

Consider the case, relevant to two phase conducting composites, where $\BL=z_1\Gc_1(\Bx)\BI+z_2\Gc_2(\Bx)\BI$. With the
choice $z_0'=\sqrt{z_1z_2}$ one has 
\beq \BK=2(\BL-z_0'\BI)(\BL+z_0'\BI)^{-1}\BL_0'=2z'_0\frac{\sqrt{z_1/z_2}-1}{\sqrt{z_1/z_2}+1}(\Gc_1-\Gc_2)\BI,
\eeq{3.8}
so that the expansion of \eq{3.1} becomes
\beqa  \BR=[\BI-\BGG_1\BB/z_0]^{-1} & = &(\BL-\BL_0')^{-1}\BK(\BI-\BGY\BK)^{-1}= (\BL-\BL_0')^{-1}\BK\sum_{n=0}^\infty(\BGY\BK)^n \nonum
& = &2\sqrt{z_1z_2}(\BL+\BI\sqrt{z_1z_2})^{-1}\sum_{n=0}^\infty [(2\Gc_1-\BI)(\BI-2\BGG_1)]^n\left[\frac{\sqrt{z_1/z_2}-1}{\sqrt{z_1/z_2}+1}\right]^n.
\eeqa{3.9}
Comparing this with \eq{0.9} we now have an expansion where effectively $\Gc_1$ and $\BGG_1$ are
replaced by $\Gc_1-\tfrac{1}{2}\BI$ and $\BGG_1-\tfrac{1}{2}\BI$ thus having the spectral radius of both. Due to the appearance of the terms $\sqrt{z_1/z_2}$
in this expansion it is best suited to the case where $\BB$ and hence $\BA$
are Hermitian. However, the expansion still works if they are not
self adjoint. For the case where $\BB(\Bx)$ takes the form
\beq \BB=z_0\BI-z_1\BP-z_2(\BI-\BP), \eeq{nsa1}
where $\BP$ is a projection, but not a Hermitian one and not necessarily
local in real space, one still has the expansion \eq{3.9} but $\BI-2\BGG_1$ no longer has
norm 1. Such expansions will be used in Section \sect{35}. 

Note that we always have the freedom to rescale a selfadjoint bounded
$\BB$ so that it is replaced by a positive semidefinite operator of
norm less than 1. To do this we rewrite the formula for $\BR$ appearing at the end of the first line in \eq{0.1a} by
\beq \BR=
[\widetilde{z}_0\BGG_1-\BGG_1\widetilde{\BB}\BGG_1]^{-1}/\Ga,\quad\text{where}
\quad \widetilde{z}_0=(z_0-c)/\Ga,\quad \widetilde{\BB} = (\BB-c\BGG_1)/\Ga,
\eeq{3.11}
in which we are free to choose $c$ and $\Ga$. Taking
\beq c=\tfrac{1}{2}(b^++b^-)\quad\text{and}\quad\Ga=\tfrac{1}{2}(b^+-b^-)
\eeq{3.12}
then guarantees that the spectrum of $\widetilde{\BB}$ is between $0$ and $1$.

We remark that this method does not always converge more rapidly than Moulinec and Suquet's method. Generally it does, with a large factor of improvement.
However, it depends on the spectrum of $\BA$, which we will consider in the next section. For example, for the conductivity of composites of two isotropic phases having
conductivities $\Gs_1$ and $\Gs_2$ , Moulinec and Suquet's method can sometimes converge for negative ratios of  $\Gs_1/\Gs_2$, this is never the case
for the ``accelerated'' method, as the square roots in \eq{3.8} induce singularities that prevent convergence when $\Gs_1/\Gs_2<0$. This is explored in more
detail in \cite{Moulinec:2018:CIM}: see \fig{ms}.

\begin{figure}[!ht]
\centering
\includegraphics[width=0.95\textwidth]{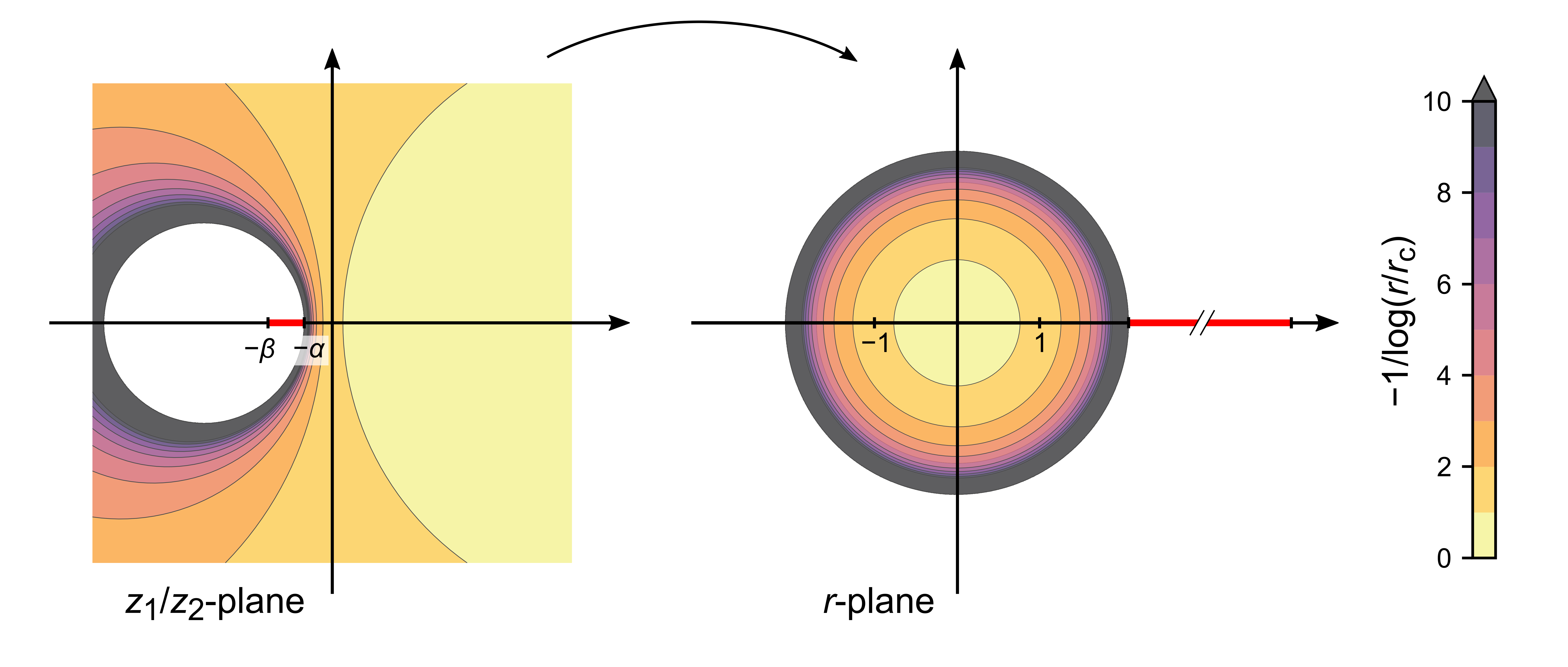}
\caption{Rates of convergence in the $z_1/z_2$-plane and in the
$r$-plane for the original
  Moulinec-Suquet scheme \protect{\cite{Moulinec:1994:FNM, Moulinec:2018:CIM}}
  where $r=(z_2-z_1)/(z_2+z_1)$.
  The intervals of possible singularities are marked
  by red lines. As concluded in \protect{\cite{Moulinec:2018:CIM}},
  even without knowledge
  of $\Ga$ and $\Gb$ (that are $\Ga=0.35$ and $\Gb=0.8$ in this example)
  their scheme can converge for negative values
  of $z_1/z_2$ and outperform the ``Eyre-Milton'' scheme in certain regions
  of the complex $z_1/z_2$-plane. The convergence rates for the
  ``Eyre-Milton'' scheme, correspond to those in
  the $\underline{z}_1/\underline{z}_2$-plane in  \protect{\fig{x}}.
  The contours reflect the number of iterations $m$ needed for
  convergence to a tolerance $\Ge$. In the $r$-plane, for small enough $\Ge$,
  one needs at radius $r$ for $m$ to be such that $c(r/r_0)^m\approx\Ge$
  for some
  constant $c$ where $r_0>1$ is the radius of convergence, i.e.,
  $m\approx \log(\Ge/c)/\log(r/r_0)$ so we have plotted the
  contours of $-1/\log(r/r_0)$ and their preimages in the
  $z_1/z_2$-plane.}
\labfig{ms}
\end{figure}

Other accelerated schemes that do not use information about  the spectrum of $\BA$ include those of Michel, Moulinec, and Suquet \cite{Michel:2000:CMB}
and Monchiet and Bonnet \cite{Monchiet:2011:PBF}. All three accelerated schemes are compared in \cite{Moulinec:2014:CTA}.

\section{Getting even faster convergence when we have bounds on the spectrum of $\BA$}
\setcounter{equation}{0}
\labsect{35}

So far in developing our expansions we have used bounds on the operator $\BB$, but using the tools presented in Section 3 of \cite{Milton:2020:UPLV}, or otherwise, we may have bounds on the
spectrum of the operator $\BA=\BGG_1\BB\BGG_1$ in the subspace $\CE$ and, as we will see now, this information can be used to obtain faster convergence.
The route explored here is by no means obvious but has its
foundations in the theory of superfunctions, including the ideas of nonorthogonal subspace collections, as
developed in Chapter 7 of \cite{Milton:2016:ETC}, and the idea of
substituting of one subspace collection into another subspace collection (first introduced in Section 29.1 of \cite{Milton:2002:TOC}). The analysis here
closely parallels that in Chapter 8 of \cite{Milton:2016:ETC}, which also outlines the reasoning for following the steps here.

To start we consider the following linear algebra problem: given $h, s, p_1,p_2,p_3$ and $E_1$, solve the matrix equation,
\beq
\bpm
J\\0 \\ J_2
\epm = z_0\bpm E\\E_1 \\ 0\epm-s\underbrace{\bpm p_1^2 & p_1p_2 & p_1p_3 \\ p_1p_2 &
p_2^2 & p_2p_3 \\  p_1p_3 & p_2p_3 & p_3^2 \epm}_{\BP}\bpm E\\E_1\\0\epm-\bpm h\\ 0 \\0\epm,
\eeq{4.1}
for $J$ in terms of $E$. We will ultimately allow for $p_1,p_2$ and $p_3$, that are either real or purely imaginary, chosen with
\beq p_3^2=1-p_1^2+p_2^2, \eeq{4.1a}
to ensure that $\BP$ is a projection matrix, though not selfadjoint in our application.
The significance of \eq{4.1} is that it corresponds to a problem in the abstract theory of composites:
define $\CU$, $\CE$, and $\CJ$ to be the three subspaces spanned by the three unit vectors
\beq \Bw_0=\bpm 1 \\0 \\ 0 \epm,\quad \Bw_1=\bpm 0 \\1 \\ 0 \epm,\quad \Bw_2=\bpm 0 \\0 \\ 1 \epm,
\eeq{4.1A}
respectively, so that $\BGG_i=\Bw_i\otimes\Bw_i$, $i=1,2,3$, are the projections onto $\CU$, $\CE$, and $\CJ$ respectively. The associated projections are
\beq \BGG_0 = \bpm 1 & 0 & 0 \\ 0 & 0 &0 \\  0 & 0 & 0 \epm,\quad
\BGG_1 = \bpm 0 & 0 & 0 \\ 0 & 1 &0 \\  0 & 0 & 0 \epm, \quad
\BGG_2 = \bpm 0 & 0 & 0 \\ 0 & 0 &0 \\  0 & 0 & 1 \epm.
\eeq{assproj}

Then \eq{4.1} reduces to
\beq J\Bw_0+ J_2\Bw_2=\BL(E\Bw_0+ E_1\Bw_1)-h\Bw_0,\quad \BL=z_0\BI-s\BP=(z_0-s)\BP+z_0(\BI-\BP),
\eeq{4.1B}
which is a problem in the abstract theory of composites, that more generally takes the form: given $\BE_0\in\CU$,
and a source term $\Bh$ in $\CH=\CU\oplus\CE\oplus\CJ$, and
an operator $\BL$ mapping $\CH$ to $\CH$, find $\BJ_0\in\CU$, $\BE\in\CE$ and $\BJ\in\CJ$ such that
\beq \BJ_0+\BJ=\BL(\BE_0+\BE)-\Bh. \eeq{etc}
In our case, the subspaces $\CU$, $\CE$, and $\CJ$ are clearly orthogonal, but $\BP$ and $\BI-\BP$ do not generally project onto orthogonal subspaces:
we have a nonorthogonal subspace collection when $p_1,p_2$ and $p_3$ are not all real.

The motivation for considering this problem is that the abstract
theory of composites applies to resistor networks with say resistors having resistances $R_1$ and $R_0$. We have the freedom to
replace every resistor in the network having resistance $R_1$ by a circuit just containing two weighted resistances 
$R_0$ and $R_2(R_1)$, where $R_2(R_1)$ is chosen so the net resistance (effective resistance) of the circuit equals $R_1$ and, say, $R_2(1)=1$.
Then the resistance $R_*(R_1,R_0)$ of the entire network as a function of $R_1$ and $R_0$
will be the same as the resistance $\widetilde{R}_*(R_2,R_0)$ of the new network, having resistances $R_0$ and $R_2$ when $R_2=R_2(R_1)$.
In particular, we can take the circuit to consist of
a weighted resistance $q_1R_0$ in series with weighted resistances $q_2R_2/t_2$ and $q_2R_0/t_0$ in parallel, where $q_1+q_2=1$ and $t_0+t_2=1$, giving
\beq R_1=q_1R_0+\frac{q_2}{t_0/R_0+t_2/R_2}. \eeq{4.1C}
Mathematically, this step of replacing every resistor in the network having resistance $R_1$ by a circuit containing the resistances $R_2(R_1)$ and
$R_0$ is an example of substitution in subspace collections. The linear algebra problem \eq{4.1} is nothing other than the equations one
solves to arrive at \eq{4.1C}, allowing for a source term $\Bs$. A field is a three dimensional vector.
The projection $\BP$ is nothing other than the projection onto the one dimensional space of fields in
the resistor $R_2$; $\CU\oplus\CE$ is the two dimensional space of fields corresponding to electrical currents,
meeting the Kirchoff condition that the net currents flowing into a node equates with the net currents flowing out
of that node, $\CU\oplus\CJ$ is the two dimensional space  of fields resulting from potential drops, $\CU$ is the one dimensional space of fields that arise in the circuit when $R_0=R_2=1$. (The spaces $\CE$ and $\CJ$ are perhaps the reverse of what one first expects, but that is because we have resistances rather than conductances).

\begin{figure}[!ht]
\centering
\includegraphics[width=0.5\textwidth]{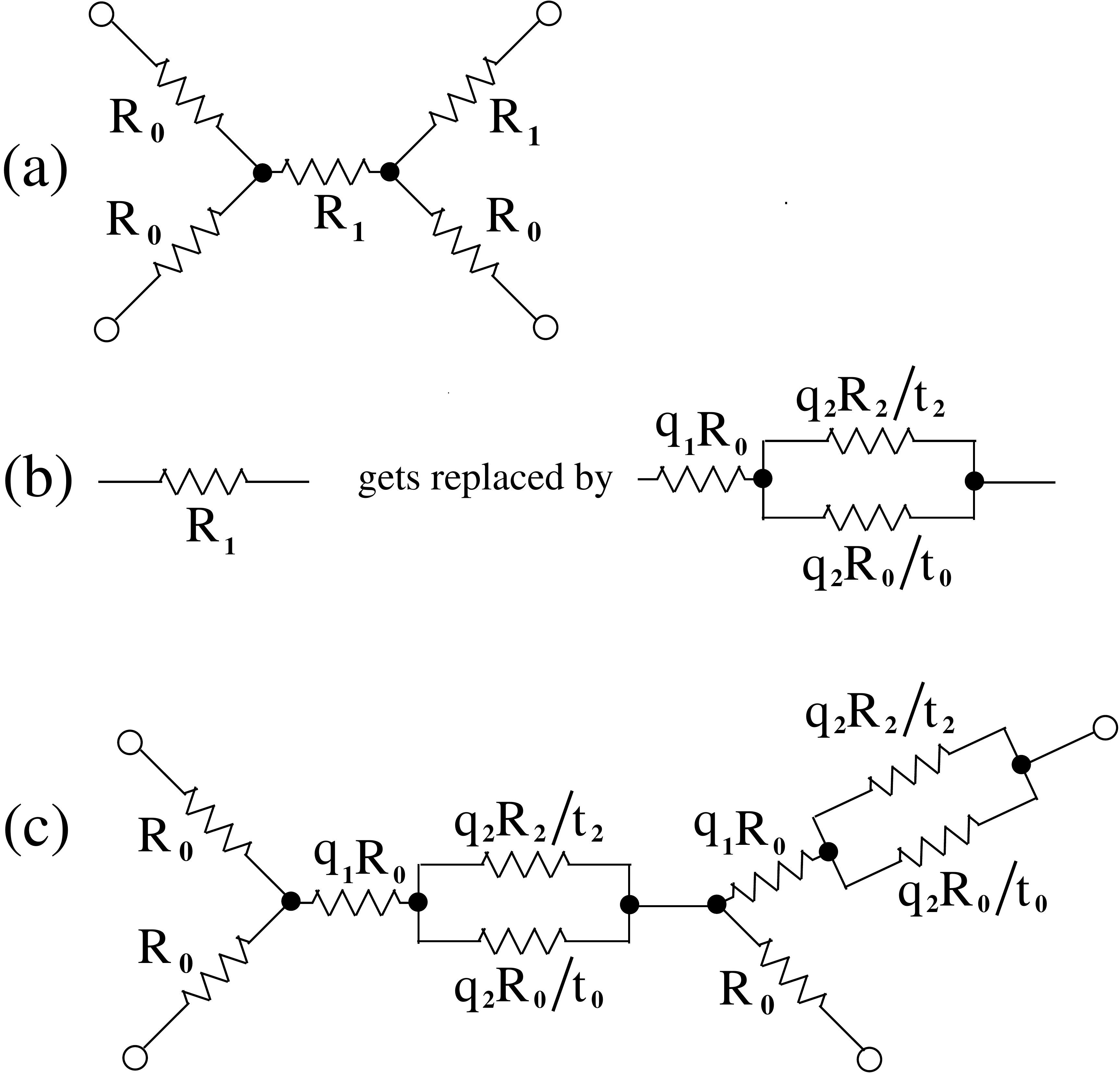}
\caption{The substitution of orthogonal subspace collections parallels that of substituting
  in a two resistor network (a), chosen to have four terminals, the subnetwork (b),
  to obtain the new network (c). If $R_2$ is chosen so the net resistance of the subnetwork
  is $R_1$ then the response of the four terminal network (c) will be the same as
  the four terminal network (a). Our substitution of nonorthogonal subspace collections
  corresponds to taking $t_0$ negative. This has a physical interpretation if we replace
  all resistors with positive resistance by capacitors and all resistors
  with negative resistance by inductors and subject the network to voltages oscillating
  with a given frequency $\Go$. Adapted from Figure 7.7 in \protect{\cite{Milton:2016:ETC}}.
  }
\labfig{subs}
\end{figure}

To find the norm of $\BP$ we consider its action on a possibly complex vector $\Ba$. We have
\beq |\BP\Ba|=|\Bp(\Bp\cdot\Ba)|\leq |\Bp|^2|\Ba|, \eeq{4.1b}
with equality when $\Ba=\Bp$. Thus $\BP$ has norm
\beq |\BP|^2=|p_1|^2+|p_2|^2+|p_3|^2, \eeq{4.1c}
and this will surely be greater than or equal to $1$ if \eq{4.1a} holds and $p_1,p_2$ and $p_3$ are either real or purely imaginary.
For example, if $p_1$ is purely imaginary while $p_2$ and $p_3$ are purely real then \eq{4.1c} implies
\beq 1=-|p_1|^2+|p_2|^2+|p_3|^2=|\BP|^2-2|p_1|^2, \eeq{4.1d}
which forces $|\BP|^2$ to be greater than or equal to $1$. 

The matrix equation \eq{4.1} is clearly satisfied with
\beqa E_1  =  \frac{sp_1p_2}{z_0-sp_2^2}E, \quad \quad
J &=& (z_0-sp_1^2)E-sp_1p_2E_1-h = \left(z_0-p_1^2s -\frac{s^2p_1^2p_2^2}{z_0-sp_2^2}\right)E-h\nonum
&=&\left(z_0-\frac{p_1^2z_0s}{z_0-sp_2^2}\right)E-h = (z_0-b)E-h,
\eeqa{4.4}
where
\beq b=\frac{p_1^2z_0}{z_0/s-p_2^2}.
\eeq{4.5}
A correspondence with \eq{4.1C} can be made by making the substitutions:
\beq s=z_0-z_2,\quad b=z_0-z_1,\quad p_2^2=t_0=1-t_2,\quad p_1^2=q_2t_2.
\eeq{equate}
Solving \eq{4.5} for $s$ in terms of $b$ gives
\beq s=\frac{bz_0}{p_1^2z_0+bp_2^2}.
\eeq{4.5a}

Suppose now that in the extended abstract theory of composites
we are interested in solving the equations
\beq \BJ(\Bx)=[z_0\BI-\BB(\Bx)]\BE(\Bx)-\Bs(\Bx),\quad\text{with}\quad \BGG_1\BE=\BE,\quad\BGG_1\BJ=0,
\eeq{4.6}
or equivalently in finding the resolvent \eq{0.1} with $\BA=\BGG_1\BB\BGG_1$.
Setting
\beq \BS(\Bx)=z_0\BB(\Bx)[p_1^2z_0\BI+p_2^2\BB(\Bx)]^{-1},
\eeq{4.7}
our preliminary linear algebra problem shows this is equivalent to solving
\beq 
\underbrace{\bpm
\BJ(\Bx)\\0 \\ \BJ_2(\Bx) \epm}_{\underline{\BJ}(\Bx)} = \Biggl[z_0\BI-\underbrace{\bpm p_1^2\BI & p_1p_2\BI & p_1p_3\BI \\ p_1p_2\BI &
p_2^2\BI & p_2p_3\BI \\  p_1p_3\BI & p_2p_3\BI & p_3^2\BI \epm
\bpm \BS(\Bx) & 0 & 0 \\ 0 & \BS(\Bx) & 0 \\ 0 &0 & \BS(\Bx)\epm}_{\underline{\BB}(\Bx)}\Biggr]\underbrace{\bpm \BE(\Bx)\\\BE_1(\Bx)\\ 0 \epm}_{\underline{\BE}(\Bx)}-
\underbrace{\bpm \Bs\\ 0 \\ 0 \epm}_{\underline{\Bs}(\Bx)},
\eeq{4.8}
with $\BGG_1\BE=\BE$ and $\BGG_1\BJ$. We are back at an equivalent problem in the extended abstract theory of composites as both $\underline{\BJ}$ and
$\underline{\BE}$ lie in orthogonal spaces. Specifically, we have
\beq \underline{\BJ}(\Bx)=\underline{\BL}(\Bx)\underline{\BE}(\Bx)
-\underline{\Bs}(\Bx),\quad
\underline{\BGG}_1\underline{\BE}=\underline{\BE},\quad \underline{\BGG}_1\underline{\BJ}=0\quad
\text{with}\quad \underline{\BL}(\Bx)=z_0\BI-\underline{\BB}(\Bx),\quad \underline{\BGG}_1=\bpm \BGG_1 & 0 & 0 \\ 0 & \BI &0 \\  0 & 0 & 0 \epm.
\eeq{4.9}
To see how this can improve convergence, let us consider the case where $\BB(\Bx)$ and $z_0$ are given by \eq{0.5a}.
Then
\beq \BS(\Bx)=\frac{z_0\Gc(\Bx)}{p_1^2z_0+p_2^2}=-\frac{z_0\Gc(\Bx)(z_1-z_2)}{p_1^2z_2-p_2^2(z_1-z_2)}, \eeq{4.10}
and associated with \eq{4.8} is the resolvent
\beq [z_0\BI-\underline{\BGG}_1\underline{\BB}]^{-1}=z_0^{-1}\{\BI-[(\underline{z}_2-\underline{z}_1)/\underline{z}_2]\underline{\BGG}_1\BGL\}^{-1}, \eeq{4.11}
where
\beq \BGL(\Bx)=\Gc(\Bx)\Bp\otimes\Bp,\quad \underline{z}_2=1,\quad
\underline{z}_1=1+ \frac{z_0(z_1-z_2)}{p_1^2z_2-p_2^2(z_1-z_2)}. \eeq{4.12}
Note that $\BGL$ is a projection operator
because both $\BP$ and $\Gc_1$ are projections and thus the operator inverse in \eq{4.11} has exactly the same form as in \eq{0.5} with
$z_1$, $z_2$ and $\Gc_1$ being replaced by $\underline{z}_1, \underline{z}_2$, and $\BGL$. Thus \eq{4.11} can be thought of as the resolvent associated with some
sort of ``two phase composite'' with moduli  $\underline{z}_1$ and $\underline{z}_2$.
Also $\underline{z}_1$ can be re-expressed in the form
\beq \underline{z}_1= 1+\frac{(z_1-z_2)(\Gb-\Ga)}{(z_1+\Gb z_2)(1+\Ga)}=\frac{(z_1+\Ga z_2)(1+\Gb)}{(z_1+\Gb z_2)(1+\Ga)},
\eeq{4.11a}
with
\beq
\Ga = -1 -\frac{p_1^2}{p_2^2-1},\quad
\Gb = -1-\frac{p_1^2}{p_2^2}.
\eeq{4.11b}
Note that $-\Ga$ (respectively $-\Gb$) is obtained by substituting $t=0$
(respectively $t=\infty$) in \eq{4.12}. Given real $\Gb>\Ga>0$ we need to
choose $p_1$ and $p_2$ so that these equations are satisfied. This will
necessitate complex solutions for $p_2$ since otherwise $\Gb$
will be negative. Explicitly we have
\beq p_1^2=\left[\frac{1}{1+\Ga}-\frac{1}{1+\Gb}\right]^{-1}, \quad  p_2^2=\left[1-\frac{1+\Gb}{1+\Ga}\right]^{-1}.
\eeq{4.12a}
With $p_1$ being real and $p_2$ being purely imaginary,
and so $\BGL$ is no longer Hermitian, even though it is a projection. This translates to a problem in the extended abstract theory of
composites with a nonorthogonal subspace collection, as introduced in Chapter 8 of \cite{Milton:2016:ETC}. 

Next, we follow the steps outlined in the previous section, though now $\BGL$ does not have norm 1. We end up with an expansion
\beqa [\BI-\BGG_1\BB/z_0]^{-1} &= & (\BL-\BL_0')^{-1}\BK(\BI-\BGY\BK)^{-1} =  (\BL-\BL_0')^{-1}\BK\sum_{n=0}^\infty(\BGY\BK)^n \nonum
& = & 2\sqrt{z_1z_2}(\BL+\BI\sqrt{z_1z_2})^{-1}\sum_{n=0}^\infty [(2\Gc_1-\BI)(\BI-2\BGG_1)]^n\left[\frac{\sqrt{z_1/z_2}-1}{\sqrt{z_1/z_2}+1}\right]^n \nonum
& = & \sum_{n=0}^\infty v^n\BC_n \nonum
& = &\sum_{n=0}^\infty \BC_n
\left(
  \frac{\sqrt{\underline{z}_1/\underline{z}_2}-1}{\sqrt{\underline{z}_1/\underline{z}_2}+1}\right)^n=
\sum_{n=0}^\infty\BC_n
\left(
\frac{\sqrt{\frac{(z_1+\Ga z_2)(1+\Gb)}{(z_1+\Gb z_2)(1+\Ga)}}-1}
{\sqrt{\frac{(z_1+\Ga z_2)(1+\Gb)}{(z_1+\Gb z_2)(1+\Ga)}}+1}
\right)^n,
\eeqa{4.13}
where
\beq \BC_n=2\sqrt{\underline{z}_1\underline{z}_2}(\underline{\BL}+\BI\sqrt{\underline{z}_1\underline{z}_2})^{-1}
\sum_{n=0}^\infty [(2\BGL-\BI)(\BI-2\underline{\BGG}_1)]^n,
\eeq{4.13a}
and
\beq  v=\frac{w-1}{w+1},\quad w=\sqrt{\underline{z}_1/\underline{z}_2},\quad \underline{z}_1/\underline{z}_2=\frac{(z_1/z_2+\Ga)(1+\Gb)}{(z_1/z_2+\Gb)(1+\Ga)}.
\eeq{4.15}
We now obtain lower bounds on the rate of convergence of the series using bounds on the spectrum of $\BA$. We suppose that the spectrum of $\BA=\BGG_1\Gc_1$ on the subspace $\CE$ lies inside the interval between $a^-$ and $a^+$ (i.e. $\BA$ satisfies \eq{0.7a})
and we let $\Ga=(1/a^-)-1$ and $\Gb=(1/a^+)-1$ so that the singularities of
$\BGG_1\BL\BGG_1$ lie between $z_1/z_2=-\Ga$ and $z_1/z_2=-\Gb$. Now $v$ is obtained from $z_1/z_2$ through
a series of transformations $z_1/z_2\rightarrow \underline{z}_1/\underline{z}_2\rightarrow w \rightarrow v$ as indicated in \fig{x}, which also
shows how the possible singularities of $\BA$ transform under these changes of variable. The mappings transform the singularities
between $z_1/z_2=-\Ga$ and $z_1/z_2=-\Gb$ in the $z_1/z_2$-plane to singularities around the edge of the unit disk in the $v$-plane.
The radius of convergence of the
series is dictated by the resolvent's nearest singularity to the origin in the $v$-plane. By construction, all singularities lie outside the unit disk
in the $v$-plane and the mapping from $\underline{z}_1/\underline{z}_2$
to $w=\sqrt{\underline{z}_1/\underline{z}_2}$ will create a singularity at the
origin in the $w$-plane corresponding to a singularity on the unit disk.
Consequently we deduce that
\beq \|(2\BGL-\BI)(\BI-2\underline{\BGG}_1)\|=1. \eeq{4.15a}
This is by no means obvious as $\BGL$, like $\BP$ in \eq{4.1}, has norm exceeding $1$. 

\begin{figure}[!ht]
\centering
\includegraphics[width=0.95\textwidth]{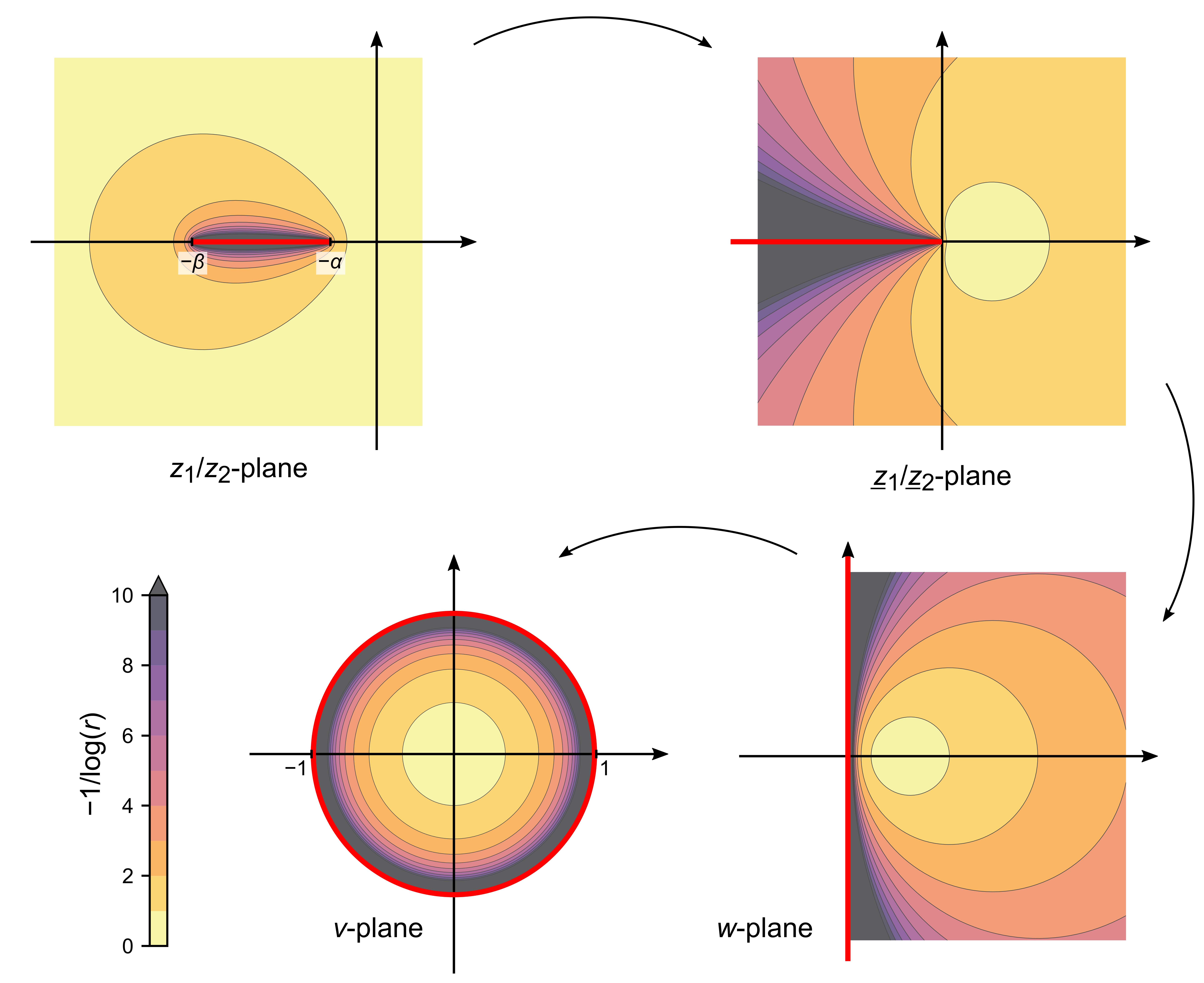}
\caption{Convergence rates in the different planes, extending
  the analysis in Chapter 8 of \protect{\cite{Milton:2016:ETC}}
  and in \protect{\cite{Moulinec:2018:CIM}}.
  The mappings transform singularities between $-\Gb$ and $-\Ga$
  (with $\Ga=0.5$ and $\Gb=2$ in this example)
  in the complex $z_1/z_2$-plane to singularities around the
  edge of the unit disk in the $v$-plane. The possible range of
  singularities are marked
  in red, though in the last two figures one could have
  singularities in the analytic continued function outside the unit disk
  in the $v$-plane or in the left hand side of the $w$-plane.
  The contours, as in \protect{\fig{ms}}, 
  reflect the number of iterations $m$ needed for
  convergence. They are level curves of  $-1/\log(r)$ in the $v$-plane
  and their preimages in the other planes. Here  $-\Gb$ and $-\Ga$ could
  be outerbounds on the spectrum, or they could be sharp bounds marking
  the endpoints of the spectrum. 
  Note that the contours in the $\underline{z}_1/\underline{z}_2$-plane
  coincide with those for the accelerated ``Eyre-Milton'' scheme in
  the $z_1/z_2$-plane, corresponding to the case $\Gb=\infty$ and $\Ga=0$.}
  \labfig{x}
\end{figure}
It is to be emphasized that $a^+$ and $a^-$ can be replaced by estimates of $a^+$ and $a^-$, such as obtained by Rayleigh Ritz methods,
or by the power method as reviewed at the beginning of Section 3 in \cite{Milton:2020:UPLV}.
One can still apply the same transformations only now $(2\BGL-\BI)(\BI-2\underline{\BGG}_1)$ will have
norm greater than $1$. 

\begin{figure}[!ht]
\centering
\includegraphics[width=0.95\textwidth]{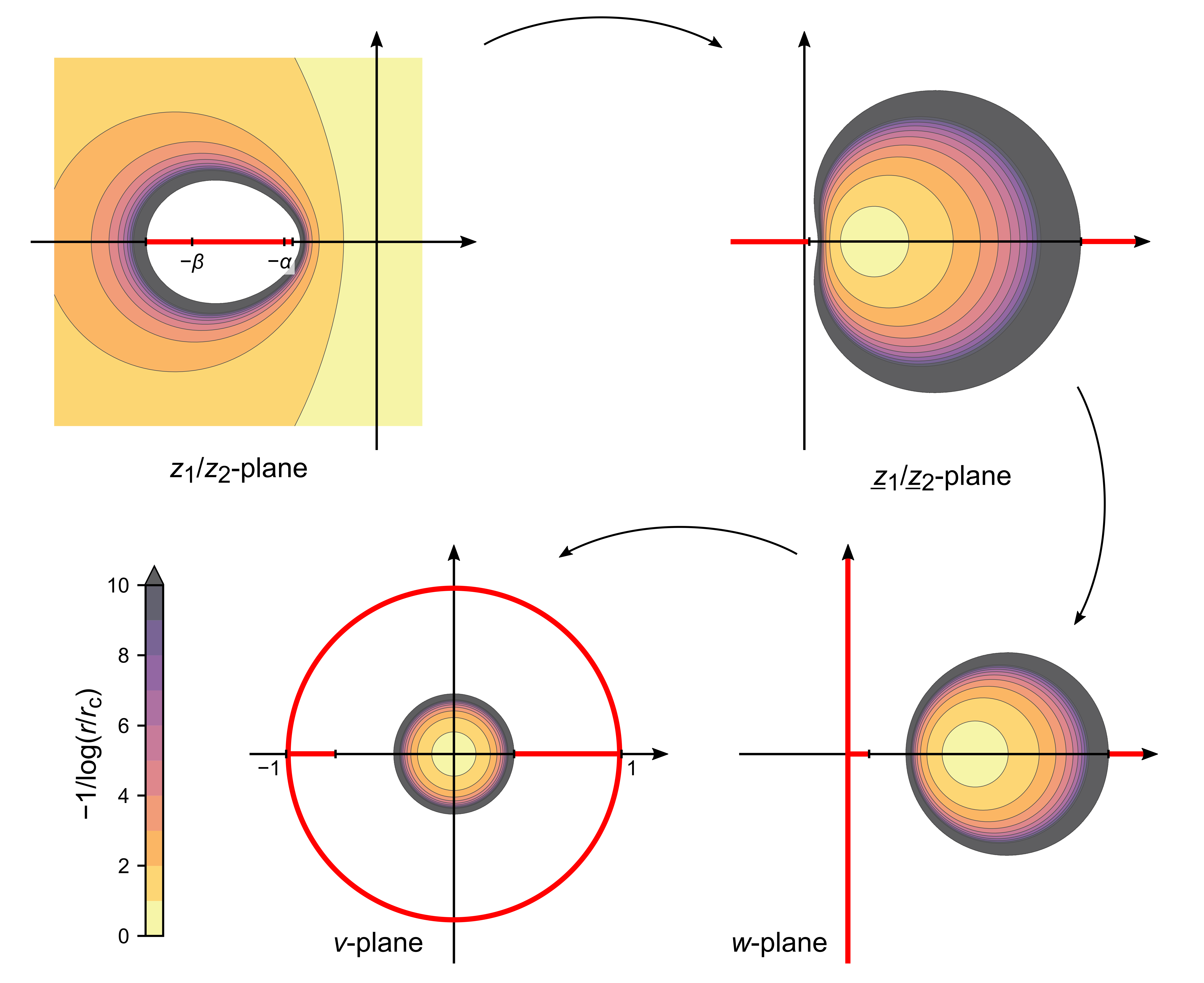}
\caption{Convergence rates when one only has estimates of $a^+$ and $a^-$
  obtained via the Rayleigh Ritz method, or by the power method. One
  can still use the same transformations. However, now there will be branch cuts extending (ideally slightly) within the unit disk in the $v$-plane,
  say a distance $d_l$ on the left side and a distance $d_r$ on the right side. As a consequence the radius of convergence $r_0<1$ of the series
  in the $v$-plane will be the minimum of $1-d_l$ and $1-d_r$, with a corresponding change in the rates of convergence of the series as
  indicated by the contours of $-1/\log(r/r_0)$ in the $v$-plane
  and their preimages in the other planes.
  As in the previous figure, the possible singularities are marked in red.
  The contours, as in \protect{\fig{ms}}, 
  reflect the number of iterations $m$ needed for
  convergence. Here  $-\Gb$ and $-\Ga$ are the
  estimates of the endpoints of the spectrum in the $z_1/z_2$ plane,
  in this example $\Ga=1$ and $\Gb=2$. The actual endpoints are the endpoints
  of the redline
}
\labfig{xx}
\end{figure}

If $\BB$ is selfadjoint but not a projection operator, it is not unclear how to choose $\Ga$ and $\Gb$ and it is also unclear how to bound the norm of the operator $\BGY\BK$.
However, after normalizing $\BB$ as in \eq{3.11} and \eq{3.12} to ensure its spectrum
is between $0$ and $1$, then it would be natural to choose $\Ga$ and $\Gb$ so that the spectrum of $\BA$
lies inside the interval between $1/(1+\Ga)$ and $1/(1+\Gb)$. To determine the success of such an approach requires further analysis and/or numerical investigations.

\section*{Acknowledgements}
GWM thanks the National Science Foundation for support
through grant DMS-1814854. The help
of Christian Kern in producing the beautiful figures is gratefully acknowledged.
\ifx \bblindex \undefined \def \bblindex #1{} \fi\ifx \bbljournal \undefined
  \def \bbljournal #1{{\em #1}\index{#1@{\em #1}}} \fi\ifx \bblnumber
  \undefined \def \bblnumber #1{{\bf #1}} \fi\ifx \bblvolume \undefined \def
  \bblvolume #1{{\bf #1}} \fi\ifx \noopsort \undefined \def \noopsort #1{}
  \fi\ifx \bblindex \undefined \def \bblindex #1{} \fi\ifx \bbljournal
  \undefined \def \bbljournal #1{{\em #1}\index{#1@{\em #1}}} \fi\ifx
  \bblnumber \undefined \def \bblnumber #1{{\bf #1}} \fi\ifx \bblvolume
  \undefined \def \bblvolume #1{{\bf #1}} \fi\ifx \noopsort \undefined \def
  \noopsort #1{} \fi

\end{document}